# Isotopically engineered silicon/silicon-germanium nanostructures as basic elements for a nuclear spin quantum computer

I Shlimak[*], V I Safarov[†] and I D Vagner[‡]

[*] Jack and Pearl Resnick Institute of Advanced Technology, Department of Physics,
Bar-Ilan University, Ramat-Gan 52900, Israel

[†] GPEC Department de Physique, Case 901 Faculté des Sciences de Luminy,
Université Aix-Marseille II 13288, Marseille, France

[‡] Grenoble High Magnetic Field Laboratory, Max Planck Institut FKF,
CRNS, F-38042, Cedex 9, Grenoble, France

E-mail: shlimai@mail.biu.ac.il

**Abstract**

The idea of quantum computation is the most promising recent developments in the high-tech domain, while experimental realization of a quantum computer poses a formidable challenge. Among the proposed models especially attractive are semiconductor based nuclear spin quantum computer's (S-NSQC), where nuclear spins are used as quantum bistable elements, "qubits", coupled to the electron spin and orbital dynamics. We propose here a scheme for implementation of basic elements for S-NSQC's which are realizable within achievements of the modern nanotechnology. These elements are expected to be based on a nuclear-spin-controlled isotopically engineered Si/SiGe heterojunction, because in these semiconductors one can vary the abundance of nuclear spins by engineering the isotopic composition. A specific device is suggested, which allows one to model the processes of recording, reading and information transfer on a quantum level using the technique of electrical detection of the magnetic state of nuclear spins. Improvement of this technique for a semiconductor system with a relatively small number of nuclei might be applied to the manipulation of nuclear spin "qubits" in the future S-NSQC.

## 1. Introduction

The idea of a quantum computer goes back to the work of Feynman [1] who pointed out, in 60-th, that classical machines can not simulate quantum systems. Indeed, the correlation between different parts of a quantum system contains exponentially large, compared to classical systems, information. Therefore, it is impossible for classical computers to simulate large quantum systems, never mind how fast they are and how much memory is available. A challenging problem is, by the other hand, to use the reach world of correlations in quantum systems in a controllable manner to process information. These would result in creation of a quantum computer. For more than two decades the problem of quantum computation has been of pure academic interest. It increased dramatically in the last few years due to discoveries of the quantum algorithms for prime factorization [2] and for exhaustive search [3] which are exponentially faster than the best known classical algorithms doing the same tasks and the sufficiently powerful quantum error-correcting codes [4]. This results in development of a few models for possible implementation of a quantum computer based on different basic quantum bistable elements - qubits (see, for example, [5-10] and references therein).

The states of spin-1/2 particles are precisely two-level systems that can be used for quantum computation. Nuclear spins are preferable because they are extremely well isolated from their environment and have a very long decoherence time. Therefore operations on nuclear spin qubits will have low error rates. Making use of the nuclear spins in semiconductors is especially promising since the modern semiconductor technology has developed extremely powerful capabilities in manipulating the charges and their spins. Current models of semiconductor based nuclear spin quantum computer (S-NSQC), could be roughly divided into two groups. One is the Kane's model [11] based on the idea of implementing a quantum computer on an array of nuclear spins located on phosphorous donors in silicon. Other approach is to incorporate nuclear spins into an electronic device (heterostructure in the quantum Hall effect regime) and to detect the spins and control their interactions electronically [12-14].

In spite of significant technological difficulties for the realization of the ingenious Kane's idea, e.g. it requires a phosphorous atoms ordered as a regular array with spacing of 20 nm as well as a gate spacing of 10 nm, his model has triggered an enormous interest to the semiconductor realizations of the nuclear spin quantum computers.

The attraction of the approach of using nuclear spin quantum computation lies in the idea of incorporating nuclear spins into a semiconductor device. One can then detect the nuclear spin polarization electronically via the hyperfine interaction between electron and nuclear spins embedded in a two-dimensional (2D) electron system in which the electron gas is in the quantum-Hall-effect (QHE) regime. As a result, nuclear spin polarization is detectable by measuring peculiarities of electron transport in a semiconductor device [15,16]. In QHE regime, at temperatures about 1 K and magnetic fields B of several Tesla, there are intervals of B where electrons fill up an integer Landau level and form a non-dissipative QHE fluid. At these fields, the Hall resistance exhibits a plateau at values of conductivity of multiples of $e^2/h$, while the dissipativity of the conduction electron gas (longitudinal magnetoresistance) approaches zero. Such systems are made usually at the interfaces between different semiconductors, the most advanced of which is the $GaAs/Al_{1-x}Ga_xAs$ heterostructure, where the highest electron mobility was achieved: $\mu > 10^7$ cm$^2$/V·s [17].

However, so far as S-NSQC is concerned, using the $GaAs/Al_{1-x}Ga_xAs$ heterostructure has very serious drawbacks. Even in the smallest lithographically prepared structures (10x10 nm), one expects a huge number of active nuclear spins ($\sim 10^4$), because all elemental components of GaAs/AlGaAs devices have non-zero nuclear spin $I$: $Ga^{69}$ ($I = 3/2$), $Ga^{71}$ ($I = 3/2$), $As^{75}$ ($I = 3/2$), $Al^{27}$ ($I = 5/2$). In this system, one cannot expect a coherent behavior of such amount of nuclear spins, it is impossible to prevent the intensive spin-flip scattering of polarized electrons on nuclear spins, which leads to a short coherent length for electrons with respect to the conservation of their spin polarization. The rate of decoherence in such a device will be too high to perform any reasonable quantum computation.

In this sense, $Si/Si_{1-x}Ge_x$ heterojunction is much more promising. Recent achievements in Si/Ge technology allow one to obtain high quality heterojunctions with mobility of about $5 \cdot 10^5$ cm$^2$/V·s [18]. Using Si/Ge heterostructures has several advantages concerning S-NSQC. First, the concentration of nuclear spins in Ge and Si crystals is much lower, because only one isotope ($Ge^{73}$ and $Si^{29}$) has a nuclear spin, and the natural abundance of this isotope is small (see Table 1). Second, the variation of isotopic composition for Ge and Si will lead to the creation of a material with a controlled concentration of nuclear spins, and even without nuclear spins. Utilization of isotopically engineered Ge and Si elements in the growth of the active Si/Ge layers, could help realize an almost zero nuclear spin layer that is coplanar with the 2DEG. Then, one might deliberately vary the isotopic composition to produce layers, wires and dots that could serve as nuclear spin qubits with a controlled number of nuclear spins. In the following section we propose the procedure how to fabricate the two-qubit basic device for S-NSQC. We do not consider on this stage the problems of a quantum register formation by an array of qubits, their coherence and how they can be brought into entanglement. In section 3, the device operation is discussed which enables to record, read and transfer information on a quantum level.

## 2. Fabrication procedure

The first step consists of the growth of isotopically engineered Si and Si-Ge epitaxial layers with lateral modulation of nuclear spin isotope content in the layer. The simplest case is a sequence of stripes of the spinless isotope (for example, $Si^{28}$) and nonzero spin isotope ($Si^{29}$, or natural Si, which contains 4.7% of $Si^{29}$, or a controlled mixture of them). The following method, based on molecular-beam-epitaxy (MBE) growth on vicinal planes, can be suggested for preparation of such a striped layer.

Usually, the substrate surface of Si cut at a small angle $\theta$ to some crystallographic direction consists of atomic size steps [19], followed by a relatively long plateau (see Fig. 1a). The typical size of the steps $a$ is of order 0.1-0.5 nm (depending on the crystallographic orientation), which gives the following value for the length of plateau $d = a/\theta = 100\text{-}500$ nm for $\theta = 10^{-3}$. The atoms deposited on the hot substrate are mobile and move toward steps in the corner where more dangling bonds are utilized (see Fig. 1b,c). As a result, during the process of deposition and formation of new layers, the steps move from left to right continuously. For a given deposition rate, one can determine the time $t$ needed for the deposition of one monolayer, i.e. to cover the entire plateau. If one isotope is deposited during some part of this time $\tau$ (Fig. 1b), followed by other isotope being deposited during the rest of this time interval $t - \tau$, then a periodical striped layer will be obtained with a controlled ratio of the strip widths: $l/(d - l) = \tau /(t- \tau)$, Fig. 1e.

The next step consists of fabricating a nanosize structure on top of this layer. For this purpose, the Scanning Probe Microscopy (SPM) technique is usually used, which can go lower than the limits obtainable by conventional methods, such as photon- and electron-beam lithography [20,21]. In our case, for fabrication of a nuclear-spin qubit device on top of an isotopically engineered Si-layer, we suggest using the Atomic Force Microscopy (AFM)-assisted local oxidation technique, which consists of the following steps.

During the nanostructure patterning, the silicon surface will be firstly H-passivated by treatment in buffered HF acid, which strips the native oxide and terminates the surface bonds with a hydride layer. This passivating hydride layer is robust and protects the surface for days from oxidation. Under suitable bias between conducting the AFM tip and the surface, the high electric field oxidizes the Si surface in the immediate vicinity of the tip. The resulting oxide pattern is transferred into the Si by selective etches, which attack silicon but not its oxide. Features down to 10 nm were obtained with this technique [22]. The steps on vicinal planes can be seen on the AFM images of the surface before patterning, and could serve as reference marks for correct alignment of nanoscale pattern with respect to the isotope $Si^{29}$ stripes.

To prepare the nanoscale structure for electrical measurements, this structure will be fabricated inside an intermediate micrometer scale pattern previously created on the top layer which is connected to the outer measurement circuit.

### 3. Device operation

The structure for S-NSQC should permit several operations: selective population with electrons of two Quantum Dots (QD), measurement of spin states in QD by the magnetoresistance, the modulation of the coupling between QD. We propose the following configuration of this structure (Fig. 2), which should be patterned on the spinless layer containing stripes of nuclei with nonzero spins. These stripes are presented as gray bands.

The pattern consists of two channels $S_1D_1$ and $S_2D_2$ (S and D denote source and drain of these channels), connected by a bridge. Applying negative voltages to the lateral gates $G_1$, $G_2$ and $G_3$ can create the quantum dots (QD) containing the nonzero spin nuclei. By proper design of the structure and by proper choice of the gate voltages, one assures that only the QD contains the nonzero spin nuclei. Applying a positive or negative voltage to the $G_3$ gates can modulate the coupling between two QD. The magnetoresistance measurements using longitudinal probes $V_1V_1$ and $V_2V_2$ will permit the detection of the spin states in the QD.

Future experiments with the above structures might proceed as follows:

1) Separate two parts of a sample shown in Fig. 2 by applying a negative voltage to the barrier gates $G_3$. Additionally, one of the dots (say, the right one) must be depleted by gate $G_2$.

2) Switch on perpendicular magnetic field which corresponds to the minimal longitudinal resistivity in the QHE regime and apply microwave radiation with frequency which correspond to the Electron Spin Resonance (ESR) in Si at given magnetic field. This should lead to intensive electron transitions with spin flipping. In its turn, this causes polarization of nuclear spins due to hyperfine interaction between electron and nuclear spin systems This step corresponds to the record of information on a quantum level.

3) Stop the microwave radiation, maintain the barrier between two parts (dots) but introduce electrons in the right part at the same density as in the left part by adjusting gate voltage $G_2$.

4) Measure the nuclear spin polarization $\eta = (n\uparrow - n\downarrow)/(n\uparrow + n\downarrow)$ in both QD using the Overhauser effect [23]. The value of $\eta$ can be detected via observation of a hysteresis in magnetoresistance measurements by sweeping magnetic field up and down, as was already demonstrated in GaAs/Al$_{1-x}$Ga$_x$As [15]. It is expected that in the left QD $\eta = 1$ while in the right QD $\eta = 0$. This step corresponds to reading information recorded on a quantum level.

5) Remove the barrier G$_3$ and allow non-dissipative motion of electrons between two parts of the structure for some fixed time interval $T$. Rebuilt the barrier again and measure $\eta$ in both parts (dots). It is expected that in the left part $\eta$ will decrease, whereas in the right part, $\eta$ will increase, reaching equal values after sufficiently long time $T$. This effect corresponds to the transfer of information recorded on a quantum level between two elementary cells (quantum dots).

Isotope engineering will allow us to dilute the number of nuclear spins in quantum dots. Therefore, implementation of the above program is also aimed to improve the technique of electrical detection of the nuclear spin state in semiconductor systems with a relatively small number of nuclei. This might be applied to the manipulation of nuclear spin qubits in the future realization of S-NSQC.

## 4. Experimental feasibility and summary.

Fabrication of this device does not present an insuperable task. Isotopically enriched compositions are available. Use of isotopically engineered Ge and Si for other goals has been reported. For example, isotopically engineered Ge$^{74}$ and Ge$^{70}$ crystals were used in a neutron-transmutation-doping technique to obtain samples with different types of conductivity and different compensation [24-26], isotopically enriched Si$^{28}$ crystals were used to increase the thermal conductivity of a Si substrate [27].

In the current state of AFM-assisted lithography, this type of structure can be fabricated with sufficiently small channel width. The main difficulty presents only the proper alignment of the pattern with respect to the isotope-engineered stripes.

The measurement of the nuclear spin polarization in quantum dots is a challenge for experimentalists, since the number of nuclear spins interacting with 2D electrons is negligibly small compared to their total number in a sample. However, successful electronic detection of the nuclear spin polarization in the QHE regime performed by von Klitzing group [15] is encouraging.

Various experimental technics were used since to measure the hyperfine coupling between the nuclear spins and the electrons in QHE regime and in mesoscopic systems [28-30]. These experiments demonstrate clearly the feasibility of measuring the polarization of a tiny amount of nuclear spins in heterojunctions via their coupling to the conduction electron spins. A novel approach in detecting nuclear spin polarization in mesoscopic systems consists in measuring the electron phase difference in nano-channels with nuclei in different states of spin polarization [31]. Hyperfine Aharonov-Bohm oscillations produced in such a way will serve as reading-off devices with a very high spatial resolution.

**Acknowledgements**

We thank Professors P Wyder, T Maniv, V Fleurov, K Kikoin, S Luryi, R G Mani and V Privman for valuable discussions.

**References**

[1] *Feynman Lectures on Computation*, Eds. Hey A.J.G. and Allen R (Perseus Press, 1996).

[2] Ekert A and Jozsa R 1996 *Rev. Mod. Phys*. **68** 733

[3] Grover L K 1997 *Phys. Rev. Lett*. **79** 325

[4] Steane A M 1996 *Phys. Rev. Lett*. **77** 793

[5] Steane A 1998 *Rep. Prog. Phys*. **61** 117

[6] Shor P W 1994 in *Proc. 35th Ann. Symp. Foundation of Computer Science* (ed. Goldwasser S), 124 (IEEE Computer Society, Los Alamitos, CA, 1994).

[7] Bennett C H 1995 *Physics Today*, **48** 24

[8] DiVincenzo D P 1995 *Science* **270** 255

[9] Bowden C M, Pethel S D 2000 *Laser Phys*. **10** 35

[10] Vrijen R *et al* 1999 quant-ph/9905096

[11] Kane B E 1998 *Nature* **393** 133

[12] Privman V, Vagner I D, Kventsel G 1998 *Phys. Lett.* A **239** 141

[13] Hansen R H, Bychkov Yu A, Vagner I D and Wyder P 1998 in *Proc. Intern. Conf. "Physical Phenomena at High Magnetic Fields-III", Tallahassee* (ed. Fisk Z, Gorkov L P and Schrieffer R J)

[14] Vagner I D and Maniv T 1988 *Phys.Rev.Lett*. **61** 1400

[15] Dobers M, Klitzing K v, Schneider J, Weimann G and Ploog K 1988 *Phys. Rev. Lett*. **61** 1650

[16] Stich B, Greulich-Weber S and Spaeth J M 1996 *Appl. Phys. Lett*. **68** 1102

[17] Stormer H 1999 *Rev. Mod. Phys*. **71** 875

[18] Ismail K, Arafa M, Saenger K L, Chu J O, and Mayaerson B S 1995 *Appl. Phys. Lett*. **66** 1077

[19] Etienne B, Laruelle F, Bloch J, Sfaxi L, Lelarge F 1995 *J. Crystal Growth*, **150** 336

[20] Dagata J A 1995 *Science* **270** 1625

[21] Marchi F *et al* 1998 *J. Vac. Scie. Technol*. B **16** 2952

[22]. Snow E S, Juan W H, Paugs S W, Cambell P M 1995 *Appl. Phys. Lett*. **66** 1729

[23] For a review, see Khutsishvili G R 1960 *Sov. Phys.-Uspekhi* **71** 285

[24] Shlimak I S *et al* 1983 *Sov. Techn. Phys. Lett*. **9** 377

[25] Rentzsch R *et al* 2000 *Phys. Stat. Sol (b)* **218** 233

[26] Watanabe M, Ootuka Y, Itoh K M, Haller E E 1998 *Phys. Rev.* B **58** 9851

[27] Capinski W S *et al* 1997 *Appl. Phys. Lett*. **71**, 2109

[28] Kane B E, Pfeiffer L N, and West K W 1992 *Phys. Rev*. B **46** 7264

[29] Wald K R *et al* 1994 *Phys. Rev. Lett.* **73** 1011

[30] Barrett S E *et al* 1994 *Phys. Rev. Lett.* **72** 1368

[31] Vagner I D *et al* 1998 *Phys. Rev. Lett.* **80** 2417

Table 1. **Isotopic composition of Si and Ge**

| Isotope | Abundance, (%) | Nuclear spin | Magnetic moment |
|---|---|---|---|
| $Si^{28}$ | 92.2 | 0 | 0 |
| $Si^{29}$ | 4.7 | 1/2 | - 0.55 |
| $Si^{30}$ | 3.1 | 0 | 0 |
| $Ge^{70}$ | 20.5 | 0 | 0 |
| $Ge^{72}$ | 27.4 | 0 | 0 |
| $Ge^{73}$ | 7.6 | 9/2 | - 0.88 |
| $Ge^{74}$ | 36.8 | 0 | 0 |
| $Ge^{76}$ | 7.7 | 0 | 0 |

**Figure captions**

Fig.1 Scheme of fabrication of a striped Si layer with a modulated concentration of nuclear spin isotopes.

Fig. 2. Scheme of a basic two-qubit device for S-NSQC, made from isotopically engineered Si/SiGe nanostructure. The strips denote the areas with non-zero nuclear spin isotopes.

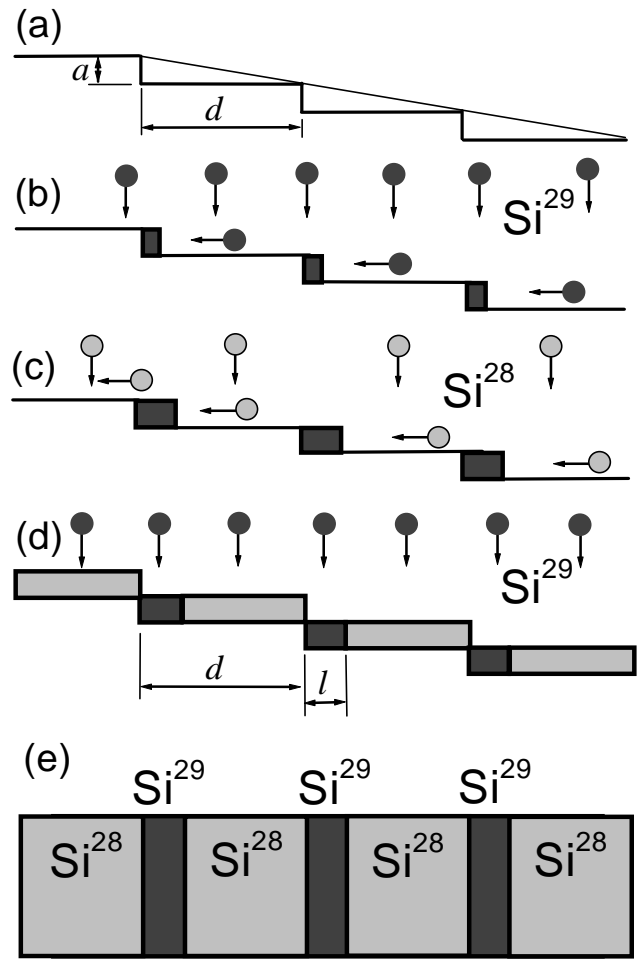

Fig. 1
Shlimak *et al*.

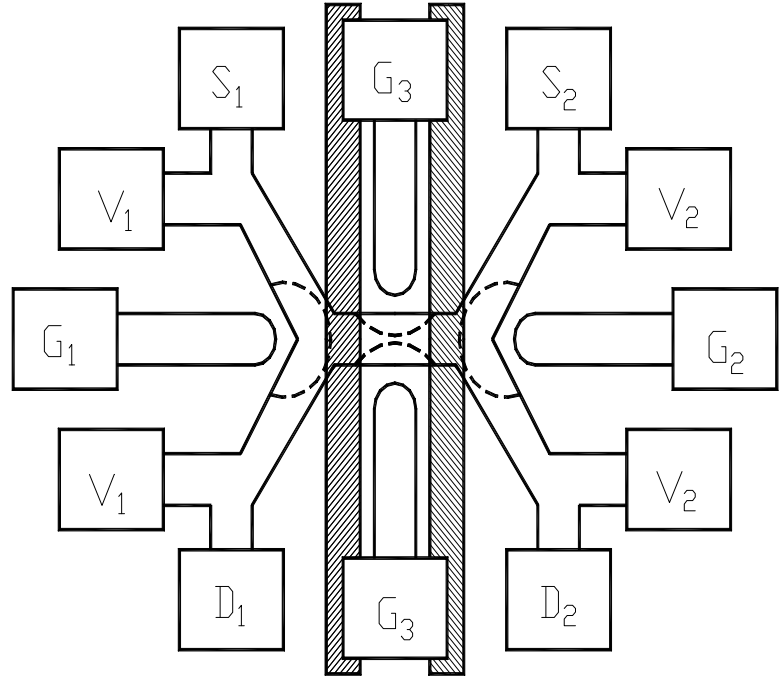

Fig.2.

Shlimak *et al*.